\PassOptionsToPackage{unicode}{hyperref}
\PassOptionsToPackage{hyphens}{url}
\documentclass[
]{article}
\usepackage{xcolor}
\usepackage{amsmath,amssymb}
\usepackage{iftex}
\ifPDFTeX
  \usepackage[T1]{fontenc}
  \usepackage[utf8]{inputenc}
  \usepackage{textcomp} 
\else 
  \usepackage{unicode-math} 
  \defaultfontfeatures{Scale=MatchLowercase}
  \defaultfontfeatures[\rmfamily]{Ligatures=TeX,Scale=1}
\fi
\usepackage{lmodern}
\ifPDFTeX\else
\fi
\IfFileExists{upquote.sty}{\usepackage{upquote}}{}
\IfFileExists{microtype.sty}{
  \usepackage[]{microtype}
  \UseMicrotypeSet[protrusion]{basicmath} 
}{}
\makeatletter
\@ifundefined{KOMAClassName}{
  \IfFileExists{parskip.sty}{%
    \usepackage{parskip}
  }{
    \setlength{\parindent}{0pt}
    \setlength{\parskip}{6pt plus 2pt minus 1pt}}
}{
  \KOMAoptions{parskip=half}}
\makeatother
\providecommand{\tightlist}{%
  \setlength{\itemsep}{0pt}\setlength{\parskip}{0pt}}
\usepackage{amssymb}
\usepackage{geometry}
\usepackage{hyperref}
\usepackage{cleveref}
\usepackage{amsmath}
\usepackage{titlesec}
\titleformat{\subsubsection}[block]{\normalfont\normalsize\bfseries}{\thesubsubsection}{1em}{}
\titlespacing*{\subsubsection}{0pt}{2ex plus 1ex minus .2ex}{1ex plus .2ex}
\usepackage{tikz}
\usetikzlibrary{arrows.meta,calc,3d}

\usepackage{bookmark}
\IfFileExists{xurl.sty}{\usepackage{xurl}}{} 
\makeatletter
\@ifundefined{xmpquote}{\newcommand{\xmpquote}[1]{#1}}{}
\makeatother
\hypersetup{
  pdftitle={From Symmetry and Continuity to the Lorentz Transformation},
  pdfauthor={Nianjun Tan, Hangzhou, China},
  pdfkeywords={\xmpquote{Special Relativity; Lorentz Transformation;
Elementary Derivation; Symmetry; Pedagogy; Principle of Relativity}},
  hidelinks,
  pdfcreator={LaTeX via pandoc}}

\title{From Symmetry and Continuity to the Lorentz Transformation}
\author{Nianjun Tan, \small Hangzhou, China}
\date{November 2, 2026}

\begin{document}
\maketitle

\section{Abstract}\label{abstract}

We derive the Lorentz transformation from spacetime symmetries and
continuity, without assuming the invariance of the speed of light.
Spacetime homogeneity yields an additive coordinate map; continuity
along four linearly independent inertial worldlines, together with
additivity, establishes linearity throughout spacetime. Spatial isotropy
and the assumed continuous local one-parameter group structure of
collinear boosts yield velocity reciprocity and determine a generalized
Lorentz-type family parameterized by \(\kappa\). The velocity law and
the empirical identification of its invariant speed with the coordinate
speed of light in vacuum select the Lorentzian branch, fixing
\(\kappa=-1/c^2\) and recovering the Lorentz transformation.

\textbf{Keywords:} Special Relativity; Lorentz Transformation;
Elementary Derivation; Symmetry; Pedagogy; Principle of Relativity

\section{1. Introduction}\label{introduction}

The Lorentz transformation is a cornerstone of special relativity, and
its derivations differ in their assumptions and mathematical methods. To
place the present work in context, we briefly review several influential
approaches.

Einstein's original derivation is based on the principle of relativity
and the constancy of the speed of light {[}1{]}. Ignatowski showed that
the general transformation law can be derived from the principle of
relativity, spacetime homogeneity, and spatial isotropy without
initially postulating the invariance of the speed of light {[}2{]}.
Subsequent derivations following this approach have differed in the
conditions used to establish linearity. Berzi and Gorini derived an
additive functional equation from spacetime homogeneity and assumed
continuity of the additive map at the origin to exclude pathological
solutions {[}3{]}. Gannett used local boundedness of the increment map
{[}4{]}, whereas Verheest used differentiability in a derivation based
on velocity transformations {[}5{]}.

Linearity alone does not guarantee velocity reciprocity. Moylan showed
that reciprocity does not follow from the principle of relativity alone
{[}6{]}.

Following the Ignatowski approach, this paper asks how continuity along
physically realizable inertial worldlines can establish linearity, and
what additional structure is needed to derive velocity reciprocity
rather than assume it. We make these assumptions explicit and examine
how experiment constrains the resulting kinematics.

\section{2. Derivation}\label{derivation}

\subsection{2.1 Assumptions}\label{assumptions}

The derivation uses the following physical assumptions.

\begin{enumerate}
\def\labelenumi{\arabic{enumi}.}
\tightlist
\item
  \textbf{Spacetime homogeneity and spatial isotropy}\\
  Spacetime homogeneity means that no spacetime point is preferred,
  while spatial isotropy means that no spatial direction is preferred.
\item
  \textbf{Principle of Relativity}\\
  The laws of physics have the same form in all inertial frames; no
  inertial frame is physically preferred.
\item
  \textbf{Continuity of particle-motion coordinates}\\
  The coordinates describing an uninterrupted particle motion are
  continuous in every inertial frame.
\item
  \textbf{Collinear group assumption}\\
  Collinear transformations between inertial frames form a continuous,
  local one-parameter group under composition.
\end{enumerate}

\subsection{2.2 Proof of Transformation
Linearity}\label{proof-of-transformation-linearity}

\subsubsection{2.2.1 Definition of the Transformation and Its
Parameters}\label{definition-of-the-transformation-and-its-parameters}

\begin{figure}[h!]
\centering
\begin{tikzpicture}[scale=1.2,>=stealth]

  \draw[->, red] (0,0) -- (5,0) node[below right] {$x'$};
  \draw[->, green!70!black] (0,0) -- (0,2.7) node[left] {$y'$};
  \draw[->, blue] (0,0) -- (0,0,4) node[above] {$z'$};
  \node[above left, blue] at (0,0) {K'};

\draw[->, very thick, red] (0,0) -- (4,0) node[below right] {$x$};
\draw[->, very thick, green!70!black] (0,0) -- (0,2) node[left] {$y$};
\draw[->, very thick, blue] (0,0) -- (0,0,3) node[above] {$z$};

\node[below left] at (-0.3,0) {K};

\draw[->,red] (3,-0.3) -- (3.5,-0.3) node[midway,below] {$\mathbf{v}$};

\filldraw (0,0) circle (1.5pt);
\node[above right] at (0,0) {O ($t=t'=0$)};

\end{tikzpicture}
\caption{Two inertial frames, $K$ and $K'$, whose origins coincide at $t=t'=0$. Frame $K'$ moves along the common $x$-axis with velocity $v$, and their spatial axes are aligned.}
\label{fig:frames-setup}
\end{figure}
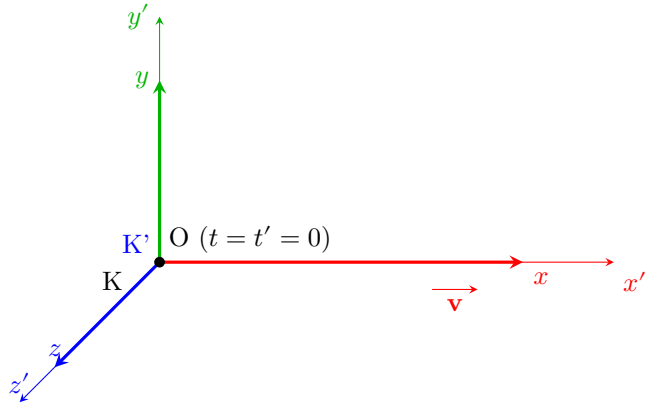

Consider two inertial reference frames, \(K(x, y, z, t)\) and
\(K'(x', y', z', t')\). Frame \(K'\) moves with constant velocity \(v\)
along their common \(x\)-axis. Their spatial and temporal origins
coincide at \(t=t'=0\), as shown in
Figure\textasciitilde{}\ref{fig:frames-setup}.

Every physical event is represented by a four-dimensional coordinate
vector,
\[\mathbf{e} = (x, y, z, t)^T \quad \text{in frame $K$,} \quad \text{and} \quad \mathbf{e}' = (x', y', z', t')^T \quad \text{in frame $K'$.}\]

Write the most general transformation between the coordinates of the
same event as \[\mathbf{e}' = T(\mathbf{e}, p)\] where \(T\) maps event
coordinates between the two frames and \(p\) denotes the parameters
describing their relative configuration.

For inertial Cartesian frames, spacetime homogeneity and spatial
isotropy allow us to choose the standard configuration described above.
In general, \(p\) may encode an origin offset, relative spatial
orientation, direction of motion, and speed. In the standard
configuration, only the signed relative velocity \(v\) along the common
\(x\)-axis remains. Thus,

\[\mathbf{e}' = T(\mathbf{e}, v)\]

\subsubsection{2.2.2 Linearity of the
Transformation}\label{linearity-of-the-transformation}

We follow Berzi and Gorini {[}3{]} in using spacetime homogeneity to
derive an additive map. Rather than assume that this map is continuous
at the origin, we invoke Assumption 3: the continuity of particle-motion
coordinates in every inertial frame. This ensures that the additive map
is continuous along inertial worldlines. We then use four physically
realizable, linearly independent inertial worldlines to establish
linearity along four independent directions; additivity extends this
linearity to all spacetime displacements.

\paragraph{\texorpdfstring{(a) Independence from the base point\\
}{(a) Independence from the base point }}\label{a-independence-from-the-base-point}

For any spacetime event \(\mathbf{e}\) and any displacement
\(\mathbf{h}\), define the increment
\[D(\mathbf{e},\mathbf{h},v):= T(\mathbf{e}+\mathbf{h},v)-T(\mathbf{e},v).\]

By spacetime homogeneity, at fixed \(v\) the image increment associated
with a displacement \(\mathbf{h}\) depends only on that displacement,
not on the starting event \(\mathbf{e}\). Equivalently, a spacetime
translation in \(K\) induces the same coordinate increment in \(K'\).
Hence \(D\) is independent of \(\mathbf{e}\), and there exists a
function \(G\) such that \begin{equation}
G(\mathbf{h},v)
= T(\mathbf{e}+\mathbf{h},v)-T(\mathbf{e},v).
\label{eq:T_increment_G}
\end{equation}

\paragraph{\texorpdfstring{(b) Reduction to additivity on full
spacetime\\
}{(b) Reduction to additivity on full spacetime }}\label{b-reduction-to-additivity-on-full-spacetime}

Let \(\mathbf{e}\) be arbitrary. Applying
Eq.\textasciitilde{}\eqref{eq:T_increment_G} twice gives \begin{align*}
G(\mathbf{h}_1+\mathbf{h}_2,v)
&= T(\mathbf{e}+\mathbf{h}_1+\mathbf{h}_2,v)-T(\mathbf{e},v) \\
&= \big[T(\mathbf{e}+\mathbf{h}_1+\mathbf{h}_2,v)-T(\mathbf{e}+\mathbf{h}_1,v)\big] + \big[T(\mathbf{e}+\mathbf{h}_1,v)-T(\mathbf{e},v)\big] \\
&= G(\mathbf{h}_2,v)+G(\mathbf{h}_1,v).
\end{align*}

Hence \(G(\cdot,v)\) is additive on the full space of displacements:
\begin{equation}
G(\mathbf{h}_1+\mathbf{h}_2,v)=G(\mathbf{h}_1,v)+G(\mathbf{h}_2,v).
\label{eq:full_additivity}
\end{equation}

\paragraph{\texorpdfstring{(c) Linearity along inertial worldlines\\
}{(c) Linearity along inertial worldlines }}\label{c-linearity-along-inertial-worldlines}

Consider a free particle in frame \(K\) that passes through an initial
event \(\mathbf{e}_{0} = (x_0, y_0, z_0, t_0)^T\) with constant
three-dimensional coordinate velocity
\(\mathbf{v} = (v_x, v_y, v_z)^T\). Its uniform rectilinear motion
produces the following spacetime displacement over a coordinate-time
interval \(t_k \in \mathbb{R}\): \begin{equation}
\mathbf{h}_k = t_k \mathbf{u}_k = (\mathbf{e}_{0} + t_k \mathbf{u}_k) - \mathbf{e}_{0}, \qquad \mathbf{u}_k := \begin{pmatrix} v_x \\ v_y \\ v_z \\ 1 \end{pmatrix},
\label{eq:inertial_displacement}
\end{equation} Here, the initial event \(\mathbf{e}_{0}\) and the
four-dimensional direction vector \(\mathbf{u}_k\) are fixed, so
\(\mathbf{h}_k\) varies continuously with \(t_k\).

We now describe the same motion in the moving frame \(K'\). By the
transformation \(T\) and the increment relation in
Eq.\textasciitilde{}\eqref{eq:T_increment_G}, its image displacement
\(\mathbf{h}_k'\) is
\[\mathbf{h}_k' = T(\mathbf{e}_{0} + t_k \mathbf{u}_k, v) - T(\mathbf{e}_{0}, v) = G(t_k \mathbf{u}_k, v) = G(\mathbf{h}_k, v).\]
We require the coordinates describing an uninterrupted particle motion
to be continuous in every inertial frame. As \(t_k\) varies continuously
in \(K\), the coordinates assigned to the corresponding events in
\(K'\), including both \(t'\) and the spatial coordinates, must also
vary continuously. A jump in a spatial coordinate would imply a finite
change in position without continuous motion. A jump in \(t'\) would
likewise assign discontinuous times to events on the same uninterrupted
worldline, even if the particle were at rest in \(K'\) and its spatial
coordinates did not jump. Assumption 3 excludes both possibilities.
Hence, \(G(t_k\mathbf{u}_k,v)\) is continuous as a function of \(t_k\).

By Eq.\textasciitilde{}\eqref{eq:full_additivity}, the restriction
\(g(t_k):=G(t_k\mathbf{u}_k,v)\) is additive in \(t_k\). The continuity
established above makes it a continuous solution of the one-dimensional
Cauchy functional equation, so it is linear. Thus, \(G\) is homogeneous
under real scaling along this inertial direction: \begin{equation}
G(\lambda\mathbf{h}_k, v) = \lambda G(\mathbf{h}_k, v), \qquad \lambda \in \mathbb{R} \quad (\text{along the inertial worldline}).
\label{eq:line_linearity}
\end{equation}

\paragraph{\texorpdfstring{(d) Extension to all spacetime
displacements\\
}{(d) Extension to all spacetime displacements }}\label{d-extension-to-all-spacetime-displacements}

Let \(\varepsilon>0\) be the numerical value of a physically realizable
nonzero coordinate speed. Choose four inertial particles with constant
three-dimensional coordinate velocities \(\mathbf{u}_0=(0,0,0)^T\) and
\(\mathbf{u}_i=\varepsilon\mathbf{e}_i\) for \(i\in\{x,y,z\}\), where
\(\mathbf{e}_i\) are the spatial unit vectors. Over a coordinate-time
interval of one unit,
Eq.\textasciitilde{}\eqref{eq:inertial_displacement} gives the following
numerical displacement vectors:
\[\mathbf{h}_0=\begin{pmatrix}0\\0\\0\\1\end{pmatrix},\quad
\mathbf{h}_x=\begin{pmatrix}\varepsilon\\0\\0\\1\end{pmatrix},\quad
\mathbf{h}_y=\begin{pmatrix}0\\\varepsilon\\0\\1\end{pmatrix},\quad
\mathbf{h}_z=\begin{pmatrix}0\\0\\\varepsilon\\1\end{pmatrix}.\]

Because \(\varepsilon\neq0\), the vectors
\(\{\mathbf{h}_0, \mathbf{h}_x, \mathbf{h}_y, \mathbf{h}_z\}\) are
linearly independent and form a basis for \(\mathbb{R}^4\). Thus any
spacetime displacement \(\mathbf{h}\)---whether or not it represents a
physical motion---can be uniquely decomposed as
\[\mathbf{h} = \alpha_0\mathbf{h}_0 + \alpha_x\mathbf{h}_x + \alpha_y\mathbf{h}_y + \alpha_z\mathbf{h}_z = \sum_{k} \alpha_k \mathbf{h}_k, \qquad k \in \{0,x,y,z\},\]
where \(\alpha_k \in \mathbb{R}\) are scalar coefficients. Using the
additivity of \(G(\cdot,v)\) on the full displacement space, established
in Eq.\textasciitilde{}\eqref{eq:full_additivity}, together with its
linearity along each basis worldline, established in
Eq.\textasciitilde{}\eqref{eq:line_linearity}, we obtain, for any real
scaling factor \(\lambda\), \begin{align*}
G(\lambda\mathbf{h},v)
&= G\left(\sum_{k}\lambda\alpha_k\mathbf{h}_k,v\right)
= \sum_{k} G(\lambda\alpha_k\mathbf{h}_k,v) \\
&= \sum_{k} \lambda\alpha_k G(\mathbf{h}_k,v)
= \lambda \sum_{k} \alpha_k G(\mathbf{h}_k,v) \\
&= \lambda G\left(\sum_{k}\alpha_k\mathbf{h}_k,v\right)
= \lambda G(\mathbf{h},v).
\end{align*} Thus, \(G(\cdot,v)\) is a linear map over \(\mathbb{R}^4\).
Hence, there is a matrix \(F(v) = [f_{ij}(v)]_{i,j=1}^4\) such that
\[G(\mathbf{h},v)=F(v)\,\mathbf{h},\]

\paragraph{\texorpdfstring{(e) Affine and linear form of \(T\)\\
}{(e) Affine and linear form of T }}\label{e-affine-and-linear-form-of-t}

Setting \(\mathbf{e}=\mathbf{0}\) in
Eq.\textasciitilde{}\eqref{eq:T_increment_G} gives
\[G(\mathbf{h},v) = T(\mathbf{h},v) - T(\mathbf{0},v),\]

and relabeling the arbitrary displacement \(\mathbf{h}\) as
\(\mathbf{e}\)
\[T(\mathbf{e},v) = G(\mathbf{e},v) + T(\mathbf{0},v) = F(v)\,\mathbf{e}+T(\mathbf{0},v),\]

so \(T\) is affine in \(\mathbf{e}\).

In the standard configuration, the origins coincide, so
\(T(\mathbf{0},v)=\mathbf{0}\). Therefore,
\[\mathbf{e}'=T(\mathbf{e},v)=F(v)\,\mathbf{e}.\]

\paragraph{\texorpdfstring{(f) Comparison with Other Approaches to
Linearity\\
}{(f) Comparison with Other Approaches to Linearity }}\label{f-comparison-with-other-approaches-to-linearity}

Once spacetime homogeneity yields an additive coordinate-increment map,
additivity alone does not guarantee linearity. For an additive map,
continuity at one point or local boundedness is sufficient to ensure
linearity. Berzi and Gorini {[}3{]} and Gannett {[}4{]} illustrate these
respective approaches, whereas Verheest {[}5{]} derives linearity using
velocity transformations and differential relations. These approaches
employ different assumptions and methods.

In the present proof, we invoke Assumption 3 only along four physically
realizable, linearly independent inertial worldlines. Continuity along
these directions establishes linearity on a basis, and additivity
extends it to all spacetime displacements. Thus, the proof requires
neither global continuity nor local boundedness or differentiability of
\(T(\,\cdot\,,v)\).

\subsection{2.3 Symmetry Constraints and Group
Structure}\label{symmetry-constraints-and-group-structure}

Having established linearity, we now determine the coefficient functions
from symmetry and the collinear group assumption.

\subsubsection{2.3.1 Setting Up the Transformation
Equations}\label{setting-up-the-transformation-equations}

In the standard configuration, let \(v\) denote the velocity of the
origin of \(K'\) measured in \(K\). By the Principle of Relativity, the
same functional family describes every boost between inertial frames. At
this stage, we write only the direct transformation,
\[\mathbf{e}' = T(\mathbf{e},v)=F(v)\,\mathbf{e},\] without identifying
the velocity parameter of the inverse transformation with \(-v\);
reciprocity will be derived below. In components, \begin{align}
x' &= f_{11}(v)x + f_{12}(v)y + f_{13}(v)z + f_{14}(v)t \label{eq:direct_x} \\
y' &= f_{21}(v)x + f_{22}(v)y + f_{23}(v)z + f_{24}(v)t \label{eq:direct_y} \\
z' &= f_{31}(v)x + f_{32}(v)y + f_{33}(v)z + f_{34}(v)t \label{eq:direct_z} \\
t' &= f_{41}(v)x + f_{42}(v)y + f_{43}(v)z + f_{44}(v)t \label{eq:direct_t}
\end{align}

\subsubsection{2.3.2 Eliminating Cross-Terms by Explicit Rotational
Comparison}\label{eliminating-cross-terms-by-explicit-rotational-comparison}

Spatial isotropy means that the choice of transverse coordinate
directions does not affect the boost to \(K'\). We therefore compare the
same boost as written in two coordinate systems in \(K\) that differ
only by a rotation about the \(x\)-axis.

\paragraph{\texorpdfstring{(a) Setup of the rotated systems\\
}{(a) Setup of the rotated systems }}\label{a-setup-of-the-rotated-systems}

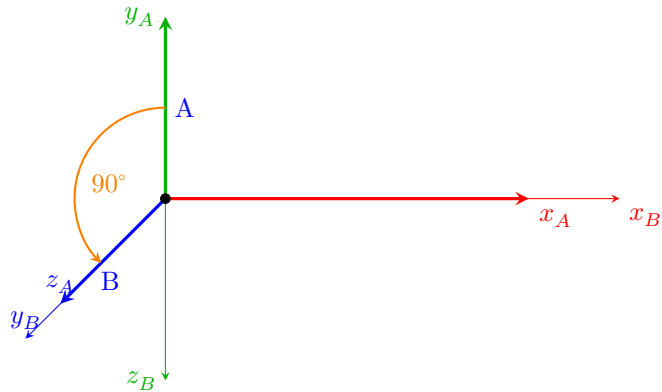
\begin{figure}[h!]
\centering
\begin{tikzpicture}[scale=1.2,>=stealth]

  \draw[->, red] (0,0) -- (5,0) node[below right] {$x_B$};
  \draw[->, blue] (0,0) -- (0,0,4) node[above] {$y_B$};
  \draw[->, green!70!black] (0,0) -- (0, -2) node[left] {$z_B$};

\draw[->, very thick, red] (0,0) -- (4,0) node[below right] {$x_A$};
\draw[->, very thick, green!70!black] (0,0) -- (0,2) node[left] {$y_A$};
\draw[->, very thick, blue] (0,0) -- (0,0,3) node[above] {$z_A$};


\filldraw (0,0) circle (1.5pt);

\draw[->,thick,orange] (0,1) arc[start angle=90, end angle=225, radius=1] node[midway, below right] {$90^\circ$};
\node[right, blue] at (0, 1) {A};
\node[left, blue] at (-0.4,-0.9) {B};

\end{tikzpicture}
\caption{Coordinate systems A and B, where B is obtained by rotating A by $90^\circ$ about the $x$-axis.}
\label{fig:Coordinate-rotation}
\end{figure}

Introduce two coordinate systems, \(A\) and \(B\), in the same inertial
frame \(K\), with \(B\) obtained by rotating \(A\) through \(90^\circ\)
about the common \(x\)-axis. Apply the same rotation in \(K'\) to obtain
\(A'\) and \(B'\). Because the same rotation is applied in both frames,
the standard configuration is preserved, so the transformations
\(A \leftrightarrow A'\) and \(B \leftrightarrow B'\) must have the same
algebraic form; see
Figure\textasciitilde{}\ref{fig:Coordinate-rotation}.

Writing the coordinates in system \(A\) as \((x_A,y_A,z_A,t_A)\), the
rotation gives
\[(x_B, y_B, z_B, t_B) = (x_A, z_A, -y_A, t_A), \quad (x'_B, y'_B, z'_B, t'_B) = (x'_A, z'_A, -y'_A, t'_A).\]

\paragraph{\texorpdfstring{(b) Comparison of coefficients\\
}{(b) Comparison of coefficients }}\label{b-comparison-of-coefficients}

Write \((x_A,y_A,z_A,t_A)\equiv(x,y,z,t)\). The transformation equations
in system \(A\) then read:
\[x'_A = f_{11}(v)x + f_{12}(v)y + f_{13}(v)z + f_{14}(v)t, \quad \text{...}\]
and in system \(B\), substituting the rotated coordinates yields:
\[x'_B = f_{11}(v)x + f_{12}(v)z - f_{13}(v)y + f_{14}(v)t, \quad \text{...}\]

Together with \((x'_B, y'_B, z'_B, t'_B) = (x'_A, z'_A, -y'_A, t'_A)\),
these relations allow us to compare coefficients for arbitrary
\((x,y,z,t)\). The \(y\)- and \(z\)-coefficients in \(x'_A\) and
\(x'_B\) give \(f_{12}(v)=f_{13}(v)=0\). Applying the same comparison to
the remaining components, particularly \(y'_B = z'_A\) and
\(z'_B = -y'_A\), yields \begin{align}
f_{12}(v) &= f_{13}(v) = f_{21}(v) = f_{31}(v) = 0 \notag\\
f_{24}(v) &= f_{34}(v) = f_{42}(v) = f_{43}(v) = 0, \notag\\
f_{22}(v) &= f_{33}(v), \quad f_{23}(v) = -f_{32}(v) \label{eq:transverse_yz}
\end{align} Thus all cross-terms between \(x\) and the transverse
coordinates vanish, and the longitudinal motion decouples from
\((y,z)\). The most general linear form compatible with these rotational
constraints is \begin{align}
x' &= f_{11}(v)\,x + f_{14}(v)\,t,  \label{eq:direct_x_2_3_2} \\
y' &= f_{22}(v)\,y + f_{23}(v)\,z,  \label{eq:direct_y_2_3_2} \\
z' &= f_{32}(v)\,y + f_{33}(v)\,z,  \label{eq:direct_z_2_3_2} \\
t' &= f_{41}(v)\,x + f_{44}(v)\,t,  \label{eq:direct_t_2_3_2}
\end{align}

\subsubsection{2.3.3 Longitudinal Symmetry and Kinematic
Parametrization}\label{longitudinal-symmetry-and-kinematic-parametrization}

Rotate the spatial axes of the standard configuration through
\(180^\circ\) about the common \(y\)-axis. Because this is a proper
spatial rotation, isotropy requires the rotated description to represent
the same physics. The rotation sends \(x\to -x\) and \(z\to -z\) while
leaving \(y\) and \(t\) unchanged. In the longitudinal \(x\)-\(t\)
sector,
\[x \to -x, \qquad t \to t, \qquad x' \to -x', \qquad t' \to t',\] and
the measured relative velocity changes sign, \(v \to -v\). No inverse
transformation is used here: coefficients evaluated at \(-v\) describe
the direct transformation after the rotation. The longitudinal
transformation \begin{align}
x' &= f_{11}(v)x + f_{14}(v)t, \\
t' &= f_{41}(v)x + f_{44}(v)t
\end{align} therefore becomes \begin{align}
-x' &= f_{11}(-v)(-x) + f_{14}(-v)t, \\
t' &= f_{41}(-v)(-x) + f_{44}(-v)t.
\end{align} Comparing the two forms for arbitrary \(x\) and \(t\) gives
\begin{align}
f_{11}(-v)&=f_{11}(v), & f_{44}(-v)&=f_{44}(v), \label{eq:diagonal_even}\\
f_{14}(-v)&=-f_{14}(v), & f_{41}(-v)&=-f_{41}(v). \label{eq:offdiagonal_odd}
\end{align} The event line of the origin of \(K'\) is \(x=vt\) in \(K\).
Substitution into the longitudinal position transformation gives
\[0=f_{11}(v)vt+f_{14}(v)t,\] and therefore \begin{equation}
f_{14}(v)=-v f_{11}(v).
\label{eq:f14_relation}
\end{equation} The longitudinal transformation is thus \begin{align}
x'&=f_{11}(v)(x-vt), \label{eq:longitudinal_x_general}\\
t'&=f_{41}(v)x+f_{44}(v)t. \label{eq:longitudinal_t_general}
\end{align} Since \(F(v)\) must be invertible, if \(f_{11}(v)=0\),
Eq.\textasciitilde{}\eqref{eq:longitudinal_x_general} would give
\(x'=0\) for every event, making the transformation noninvertible.
Therefore, \begin{equation}
f_{11}(v)\neq0
\label{eq:f11_nonzero}
\end{equation} for every allowed \(v\).

\subsubsection{2.3.4 Commutativity and Velocity
Reciprocity}\label{commutativity-and-velocity-reciprocity}

\paragraph{\texorpdfstring{(a) Commutativity of collinear boosts\\
}{(a) Commutativity of collinear boosts }}\label{a-commutativity-of-collinear-boosts}

By assumption, collinear transformations form a continuous local
one-parameter group {[}7{]}. Such a group is commutative wherever
composition is defined: \begin{equation}
F(u)F(v)=F(v)F(u).
\label{eq:boost_commutativity}
\end{equation} Because the transverse and longitudinal sectors are
decoupled, it suffices to work in the \(x\)-\(t\) sector. A boost of
velocity \(v\) acts on \((x,t)^T\) through \[F_{xt}(v)=
\begin{pmatrix}
f_{11}(v)&-v f_{11}(v)\\
f_{41}(v)&f_{44}(v)
\end{pmatrix}.\] Applying first a boost of velocity \(v\) and then one
of velocity \(u\) gives \(F_{xt}(u)F_{xt}(v)\); reversing the order
gives \(F_{xt}(v)F_{xt}(u)\). The two products are \begin{align}
F_{xt}(u)F_{xt}(v)
&=
\begin{pmatrix}
f_{11}(u)f_{11}(v)-u f_{11}(u)f_{41}(v)&-v f_{11}(u)f_{11}(v)-u f_{11}(u)f_{44}(v)\\
f_{41}(u)f_{11}(v)+f_{44}(u)f_{41}(v)&-v f_{41}(u)f_{11}(v)+f_{44}(u)f_{44}(v)
\end{pmatrix}, \label{eq:F_u_F_v}\\
F_{xt}(v)F_{xt}(u)
&=
\begin{pmatrix}
f_{11}(v)f_{11}(u)-v f_{11}(v)f_{41}(u)&-u f_{11}(v)f_{11}(u)-v f_{11}(v)f_{44}(u)\\
f_{41}(v)f_{11}(u)+f_{44}(v)f_{41}(u)&-u f_{41}(v)f_{11}(u)+f_{44}(v)f_{44}(u)
\end{pmatrix}. \label{eq:F_v_F_u}
\end{align}

\paragraph{\texorpdfstring{(b) Equality of the longitudinal diagonal
coefficients\\
}{(b) Equality of the longitudinal diagonal coefficients }}\label{b-equality-of-the-longitudinal-diagonal-coefficients}

Equating the upper-right elements of
Eqs.\textasciitilde{}\eqref{eq:F_u_F_v} and \eqref{eq:F_v_F_u}, and
dividing by the nonzero product \(f_{11}(u)f_{11}(v)\), gives
\begin{equation}
u-v=u\tau(v)-v\tau(u),
\qquad
  \tau(v):=\frac{f_{44}(v)}{f_{11}(v)}.
\label{eq:tau_relation}
\end{equation} Eqs.\textasciitilde{}\eqref{eq:diagonal_even} imply
\(\tau(-v)=\tau(v)\). Setting \(u=-v\) in
Eq.\textasciitilde{}\eqref{eq:tau_relation} therefore gives, for
\(v\neq0\), \[-2v=-2v\tau(v),\] so \(\tau(v)=1\). At \(v=0\), the
identity transformation gives \(f_{11}(0)=f_{44}(0)=1\). Hence
\begin{equation}
f_{44}(v)=f_{11}(v),\qquad \text{throughout the local velocity domain.}
\label{eq:f44_equals_f11}
\end{equation}

\paragraph{\texorpdfstring{(c) Derivation of velocity reciprocity\\
}{(c) Derivation of velocity reciprocity }}\label{c-derivation-of-velocity-reciprocity}

Let \(\bar v\) denote the velocity of the origin of \(K\) as measured in
\(K'\). The origin of \(K\) has event line \(x=0\), for which
Eqs.\textasciitilde{}\eqref{eq:longitudinal_x_general},
\eqref{eq:longitudinal_t_general}, and \eqref{eq:f44_equals_f11} give
\[x'=-v f_{11}(v)t,
\qquad
t'=f_{11}(v)t.\] It follows directly that \begin{equation}
\bar v=\left.\frac{dx'}{dt'}\right|_{x=0}=-v.
\label{eq:velocity_reciprocity}
\end{equation} Thus velocity reciprocity is derived rather than
postulated. Since the group inverse represents the transformation from
\(K'\) to \(K\), Eq.\textasciitilde{}\eqref{eq:velocity_reciprocity} now
justifies \begin{equation}
F(v)^{-1}=F(-v).
\label{eq:inverse_after_reciprocity}
\end{equation} This result agrees with the reciprocity theorem of Berzi
and Gorini {[}3{]} and identifies, within the present elementary
framework, the assumptions that supplement the Principle of Relativity
{[}6{]}.

\paragraph{\texorpdfstring{(d) Universal off-diagonal coefficient\\
}{(d) Universal off-diagonal coefficient }}\label{d-universal-off-diagonal-coefficient}

Equating the upper-left elements of
Eqs.\textasciitilde{}\eqref{eq:F_u_F_v} and \eqref{eq:F_v_F_u} gives the
second independent consequence of commutativity: \begin{equation}
u f_{11}(u)f_{41}(v)=v f_{11}(v)f_{41}(u).
\label{eq:f11_f41_cross_relation}
\end{equation} For nonzero \(u\) and \(v\),
Eq.\textasciitilde{}\eqref{eq:f11_nonzero} allows both sides to be
divided by \(uv\,f_{11}(u)f_{11}(v)\):
\[\frac{f_{41}(v)}{v f_{11}(v)}=\frac{f_{41}(u)}{u f_{11}(u)}.\] Since
the left-hand side depends only on \(v\) and the right-hand side only on
\(u\), this ratio is constant on the connected local velocity domain.
For \(v\neq0\), define \[\kappa:=\frac{f_{41}(v)}{v f_{11}(v)}.\] It
follows that, for every nonzero \(v\) in the local velocity domain,
\begin{equation}
f_{41}(v)=\kappa v f_{11}(v)
\label{eq:f41_kappa}
\end{equation} Setting \(v=0\) in
Eq.\textasciitilde{}\eqref{eq:f11_f41_cross_relation} gives
\(f_{41}(0)=0\), so Eq.\textasciitilde{}\eqref{eq:f41_kappa} holds at
\(v=0\) as well and therefore throughout the local velocity domain.

\subsubsection{2.3.5 Completing the Generalized
Transformation}\label{completing-the-generalized-transformation}

\paragraph{\texorpdfstring{(a) Transverse coefficients\\
}{(a) Transverse coefficients }}\label{a-transverse-coefficients}

To determine the transverse coefficients, consider the event
\((x,y,z,t)=(0,y,0,0)\) with \(y>0\).
Eq.\textasciitilde{}\eqref{eq:direct_t_2_3_2} gives \(t'=0\). At
\(t=t'=0\) in the standard configuration, the coordinate axes of \(K\)
and \(K'\) coincide, so this event lies on the common positive \(y\)-
and \(y'\)-axes; hence \(z=0\) and \(z'=0\).
Eq.\textasciitilde{}\eqref{eq:direct_z_2_3_2} gives \[z'=f_{32}(v)y.\]
Since \(z'=0\) and \(y>0\), we obtain \(f_{32}(v)=0\).
Eq.\textasciitilde{}\eqref{eq:transverse_yz} then gives \(f_{23}(v)=0\).
Substituting into Eq.\textasciitilde{}\eqref{eq:direct_y_2_3_2} yields
\(y'=f_{22}(v)y\); since \(y'>0\) and \(y>0\), it follows that
\(f_{22}(v)>0\).
Eq.\textasciitilde{}\eqref{eq:inverse_after_reciprocity} gives
\[f_{22}(v)f_{22}(-v)=1.\] Under the \(180^\circ\) rotation about the
\(y\)-axis used above, \(v\) changes to \(-v\) while the
\(y\)-coordinate is unchanged. Spatial isotropy therefore gives
\(f_{22}(-v)=f_{22}(v)\), so \(f_{22}(v)=\pm1\). With \(f_{22}(v)>0\)
from above, \(f_{22}(v)=1\), and
Eq.\textasciitilde{}\eqref{eq:transverse_yz} gives \begin{equation}
f_{22}(v)=f_{33}(v)=1.
\label{eq:transverse_coefficients}
\end{equation}

\paragraph{\texorpdfstring{(b) Determination of \(f_{11}(v)\)\\
}{(b) Determination of f\_\{11\}(v) }}\label{b-determination-of-f_11v}

Combining Eqs.\textasciitilde{}\eqref{eq:f44_equals_f11},
\eqref{eq:f41_kappa}, and \eqref{eq:f14_relation}, the longitudinal
block becomes \begin{equation}
F_{xt}(v)=f_{11}(v)
\begin{pmatrix}
1&-v\\
\kappa v&1
\end{pmatrix}.
\label{eq:F_xt_kappa}
\end{equation} By
Eq.\textasciitilde{}\eqref{eq:inverse_after_reciprocity},
\(F_{xt}(-v)F_{xt}(v)=I\), which gives \begin{equation}
f_{11}(v)^2(1+\kappa v^2)=1.
\label{eq:inverse_consistency_kappa}
\end{equation} The coefficient \(f_{11}(v)\) varies continuously on the
connected local velocity domain containing the identity and satisfies
\(f_{11}(0)=1\). Since Eq.\textasciitilde{}\eqref{eq:f11_nonzero} shows
that it never vanishes, it cannot change sign on that domain. Hence
\(f_{11}(v)>0\), and the positive square-root branch must be selected:
\begin{equation}
f_{11}(v)=\frac{1}{\sqrt{1+\kappa v^2}}.
\label{eq:f11_kappa}
\end{equation}

\paragraph{\texorpdfstring{(c) Relative-velocity composition law\\
}{(c) Relative-velocity composition law }}\label{c-relative-velocity-composition-law}

Suppose a boost of velocity \(v\) is followed by one of velocity \(u\),
with both velocities lying in the local velocity domain where the
composition is defined. Let \(w(u,v)\) denote the physical relative
velocity of the composite boost, defined by \(F(w)=F(u)F(v)\). Since the
transverse coefficients are unity by
Eq.\textasciitilde{}\eqref{eq:transverse_coefficients}, it is enough to
impose \(F_{xt}(w)=F_{xt}(u)F_{xt}(v)\). From
Eq.\textasciitilde{}\eqref{eq:F_xt_kappa}, \begin{equation}
F_{xt}(u)F_{xt}(v)
=f_{11}(u)f_{11}(v)
\begin{pmatrix}
1&-u\\
\kappa u&1
\end{pmatrix}
\begin{pmatrix}
1&-v\\
\kappa v&1
\end{pmatrix}
=f_{11}(u)f_{11}(v)
\begin{pmatrix}
1-\kappa uv&-(u+v)\\
\kappa(u+v)&1-\kappa uv
\end{pmatrix}.
\label{eq:F_xt_product}
\end{equation} On the other hand,
Eq.\textasciitilde{}\eqref{eq:F_xt_kappa} gives \[F_{xt}(w)=f_{11}(w)
\begin{pmatrix}
1&-w\\
\kappa w&1
\end{pmatrix}.\] Equating the \((1,1)\) and \((1,2)\) elements of
Eqs.\textasciitilde{}\eqref{eq:F_xt_product} and this expression yields
\begin{align}
f_{11}(w)&=f_{11}(u)f_{11}(v)(1-\kappa uv), \label{eq:composition_f11_w}\\
w f_{11}(w)&=f_{11}(u)f_{11}(v)(u+v). \label{eq:composition_w_f11w}
\end{align} Dividing Eq.\textasciitilde{}\eqref{eq:composition_w_f11w}
by Eq.\textasciitilde{}\eqref{eq:composition_f11_w}, with
\(f_{11}(w)\neq0\), gives \begin{equation}
w=\frac{u+v}{1-\kappa uv}.
\label{eq:combined_velocity_kappa}
\end{equation} This is the local composition law for physical relative
velocities throughout the local velocity domain, wherever composition is
defined.

\paragraph{\texorpdfstring{(d) Generalized spacetime transformation\\
}{(d) Generalized spacetime transformation }}\label{d-generalized-spacetime-transformation}

Define \begin{equation}
\gamma_\kappa(v):=\frac{1}{\sqrt{1+\kappa v^2}},
\label{eq:gamma_kappa}
\end{equation} which coincides with \(f_{11}(v)\) from
Eq.\textasciitilde{}\eqref{eq:f11_kappa}. Collecting the longitudinal
and transverse results from above, the generalized inertial-frame
transformation is \begin{align}
x' &= \gamma_\kappa(v)(x-vt), \label{eq:direct_x_2_3_6}\\
t' &= \gamma_\kappa(v)\left(t+\kappa v x\right), \label{eq:direct_t_2_3_6}\\
y'&=y,\qquad z'=z.
\end{align} By Eq.\textasciitilde{}\eqref{eq:inverse_after_reciprocity},
\(F_{xt}(v)^{-1}=F_{xt}(-v)\); in component form, \begin{align}
x &= \gamma_\kappa(v)(x'+vt'), \label{eq:inverse_x_2_3_6}\\
t &= \gamma_\kappa(v)\left(t'-\kappa v x'\right). \label{eq:inverse_t_2_3_6}
\end{align}

\subsection{2.4 Invariant Speed, Empirical Selection, and the Lorentz
Transformation}\label{invariant-speed-empirical-selection-and-the-lorentz-transformation}

The preceding analysis determines the kinematics up to the universal
constant \(\kappa\). We now derive the collinear particle-speed law,
classify the resulting branches, select the physical branch empirically,
and verify the consistency of the three-dimensional formulation.

\subsubsection{2.4.1 Collinear Velocity
Transformation}\label{collinear-velocity-transformation}

Let \(K'\) move with velocity \(v\) relative to \(K\) along the positive
\(x\)-direction. A particle or signal moving collinearly with speed
\(u\) in \(K\) has event line \(x=ut\). Applying
Eqs.\textasciitilde{}\eqref{eq:direct_x_2_3_6} and
\eqref{eq:direct_t_2_3_6} to the event line gives
\[x' = \gamma_\kappa(v)(u-v)t, \qquad t' = \gamma_\kappa(v)\left(1+\kappa uv\right)t.\]
Provided that \(1+\kappa uv\neq0\), the transformed collinear speed is
\begin{equation}
u' = \frac{x'}{t'} = \frac{u-v}{1+\kappa uv}. \label{eq:collinear_velocity}
\end{equation}

\subsubsection{2.4.2 Kinematic Branches and the Invariant
Speed}\label{kinematic-branches-and-the-invariant-speed}

We now use the collinear velocity law to determine whether each branch
admits a unique finite boost-invariant speed magnitude within the domain
of the local transformation.

\textbf{The case \(\kappa=0\).} For \(\kappa=0\),
Eq.\textasciitilde{}\eqref{eq:collinear_velocity} gives \(u'=u-v\), the
Galilean velocity-addition law, and the spacetime transformation reduces
to \(x'=x-vt\) and \(t'=t\). For \(v\neq0\),
Eq.\textasciitilde{}\eqref{eq:collinear_velocity} with \(u'=u\) gives no
finite invariant speed.

\textbf{The \(\kappa>0\) local branch.} For \(v\neq0\),
Eq.\textasciitilde{}\eqref{eq:collinear_velocity} with \(u'=u\) gives
\(\kappa u^2=-1\). Because \(\kappa>0\), no real finite invariant speed
exists.

\textbf{The \(\kappa<0\) branch.} When \(\kappa<0\), define
\[c \equiv \frac{1}{\sqrt{-\kappa}}>0.\] Since \(\kappa<0\),
Eq.\textasciitilde{}\eqref{eq:gamma_kappa} is real only when
\(1+\kappa v^2>0\), i.e.~\(|v|<c\).
Eq.\textasciitilde{}\eqref{eq:collinear_velocity} then becomes
\[u'=\frac{u-v}{1-\frac{uv}{c^2}}.\] For a nontrivial boost, the
invariance condition \(u'=u\) gives \(u^2=c^2\), and hence \(u=\pm c\).
The two signs represent opposite directions along the \(x\)-axis and
therefore define a single invariant speed magnitude \(c>0\). There are
no other real solutions.

Thus, only the \(\kappa<0\) branch admits a unique finite
boost-invariant speed magnitude, \(c=1/\sqrt{-\kappa}\), in the
collinear setting.

\subsubsection{2.4.3 Empirical Selection of the Lorentz
Branch}\label{empirical-selection-of-the-lorentz-branch}

Symmetry alone leaves \(\kappa\) undetermined: it neither selects among
the three kinematic branches nor fixes the value of any invariant speed.
We therefore use the empirical identification of the invariant speed
with the vacuum speed of light, \(c\). In the Einstein synchronization
convention used throughout this article, this means that light in vacuum
has the same coordinate speed in every inertial frame. Michelson--Morley
experiments test the isotropy of the round-trip speed of light {[}8{]},
while Kennedy--Thorndike-type, cavity, and Compton-edge experiments test
boost invariance {[}9--11{]}.

Combining this empirical input with the classification above selects the
\(\kappa<0\) branch and excludes the \(\kappa=0\) and \(\kappa>0\)
branches. Identifying the resulting invariant speed,
\(c=1/\sqrt{-\kappa}\), with the vacuum speed of light fixes
\[\kappa=-\frac{1}{c^2}<0\] and selects the Lorentzian branch. The
relative-velocity composition law,
Eq.\textasciitilde{}\eqref{eq:combined_velocity_kappa}, becomes
\begin{equation}
w = \frac{u+v}{1+\frac{uv}{c^2}}. \label{eq:combined_velocity_c}
\end{equation}

\subsubsection{2.4.4 The Lorentz
Transformation}\label{the-lorentz-transformation}

With \(\kappa=-1/c^2\) now fixed,
Eqs.\textasciitilde{}\eqref{eq:direct_x_2_3_6} and
\eqref{eq:direct_t_2_3_6} reduce to the Lorentz transformation:
\begin{align}
x' &= \gamma(v) (x - v t), \label{eq:LT_x} \\
t' &= \gamma(v) \left( t - \frac{v}{c^2} x \right), \label{eq:LT_t}
\end{align} with \begin{equation}
\gamma(v) = \frac{1}{\sqrt{1 - \frac{v^2}{c^2}}}, \qquad |v|<c. \label{eq:gamma}
\end{equation} Together with \(y'=y\) and \(z'=z\), these equations give
the Lorentz transformation in standard configuration. The inverse
transformation follows by replacing \(v\) with \(-v\). A boost in any
direction can be brought to this standard configuration by rotating both
frames.

\subsubsection{2.4.5 Three-Dimensional Velocity Law and Consistency with
Observation}\label{three-dimensional-velocity-law-and-consistency-with-observation}

Differentiate the direct Lorentz transformation along a particle
worldline, where \(u_x=dx/dt\), \(u_y=dy/dt\), and \(u_z=dz/dt\). Since
\(y'=y\), the transverse displacement satisfies \(dy'=dy=u_y\,dt\),
while
\[dt' = \gamma(v)\left(dt-\frac{v}{c^2}dx\right)=\gamma(v)\left(1-\frac{u_xv}{c^2}\right)dt.\]
Proceeding in the same way for the longitudinal and transverse
components gives \[u_x' = \frac{u_x-v}{1-\frac{u_x v}{c^2}}, \qquad
u_y' = \frac{u_y}{\gamma(v)\left(1-\frac{u_x v}{c^2}\right)}, \qquad
u_z' = \frac{u_z}{\gamma(v)\left(1-\frac{u_x v}{c^2}\right)}.\] For a
particle whose speed in \(K\) is \(u\), with \(u^2=u_x^2+u_y^2+u_z^2\),
these formulas give \begin{align*}
c^2-u'^2
&= c^2-(u_x'^2+u_y'^2+u_z'^2) = \frac{c^2-u^2}{\gamma(v)^2\left(1-\frac{u_x v}{c^2}\right)^2}.
\end{align*} Consequently, if \(u<c\) in one inertial frame, then
\(u'<c\) in every other inertial frame; if \(u=c\), then \(u'=c\). The
branch selected by the collinear analysis is therefore consistent with
the three-dimensional velocity law, and \(c\) is a boost-invariant
coordinate-speed magnitude in every direction. Under the synchronization
convention adopted above, electromagnetic radiation in vacuum has this
coordinate speed.

\section{3. Conclusion}\label{conclusion}

Starting from symmetry and worldline continuity, the argument obtains
additivity, linearity, and velocity reciprocity step by step, and only
then uses experiment to select the Lorentzian branch from the space of
possible kinematics. This separation of symmetry postulates and
empirical input offers a clean pedagogical route to special relativity.

\section{References}\label{references}

{[}1{]} A. Einstein, ``Zur Elektrodynamik bewegter Körper,'' Ann. Phys.
17, 891--921 (1905).\\
{[}2{]} W. von Ignatowski, ``Einige allgemeine Bemerkungen zum
Relativitätsprinzip,'' Phys. Z. 11, 972--976 (1910).\\
{[}3{]} V. Berzi and V. Gorini, ``Reciprocity Principle and the Lorentz
Transformations,'' J. Math. Phys. 10, 1518--1524 (1969).\\
{[}4{]} J. W. Gannett, ``Nothing but Relativity, Redux,'' Eur. J. Phys.
28, 1145--1150 (2007).\\
{[}5{]} F. Verheest, ``On the linearity of the generalized Lorentz
transformation,'' Am. J. Phys. 90, 425--429 (2022).\\
{[}6{]} P. Moylan, ``Velocity reciprocity and the relativity
principle,'' Am. J. Phys. 90, 126--134 (2022).\\
{[}7{]} B. C. Hall,
\textit{Lie Groups, Lie Algebras, and Representations: An Elementary Introduction},
2nd ed.~(Springer, 2015).\\
{[}8{]} A. A. Michelson and E. W. Morley, ``On the Relative Motion of
the Earth and the Luminiferous Ether,'' Am. J. Sci. 34, 333--345
(1887).\\
{[}9{]} R. J. Kennedy and E. M. Thorndike, ``Experimental establishment
of the relativity of time,'' Phys. Rev.~42, 400--418 (1932).\\
{[}10{]} M. Nagel et al., ``Direct terrestrial test of Lorentz symmetry
in electrodynamics to \(10^{-18}\),'' Nat. Commun. 6, 8174 (2015).\\
{[}11{]} V. G. Gurzadyan and A. T. Margaryan, ``The light speed versus
the observer: the Kennedy--Thorndike test from GRAAL-ESRF,'' Eur. Phys.
J. C 78, 607 (2018).

\end{document}